\documentclass[superscriptaddress,secnumarabic,amssymb,amsmath,nobibnotes,aps,prd,nofootinbib,preprint]{revtex4}
\usepackage{graphicx}
\usepackage{float}
\usepackage{bm}
\usepackage{amsmath}
\usepackage{amsfonts}
\usepackage{amssymb}
\usepackage{epstopdf}
\usepackage{natbib}%
\newcommand{\bee}{\begin{equation}}
\newcommand{\eee}{\end{equation}}
\newcommand{\eaa}{\end{eqnarray}}
\newcommand{\baa}{\begin{eqnarray}}

\usepackage{color}

\begin{document}
\title{Generalized Entropies and Black Hole Area Quantization from Landauer's Principle}                          

\author{Jorge Ananias Neto}
\email{jorge.ananias@ufjf.br}
\affiliation{Departamento de F\'isica, Universidade Federal de Juiz de Fora, Juiz de Fora, MG, Brazil}
\author{Ronaldo Thibes}
\email{thibes@uesb.edu.br}
\affiliation{Universidade Estadual do Sudoeste da Bahia, DCEN, Itapetinga, BA, Brazil}

\begin{abstract}
We investigate black hole area quantization by imposing Landauer's principle on the discrete entropy change between consecutive area levels. 
The elementary transition is identified with the entropy cost of erasing one bit of information, \(\Delta S=k_B\ln 2\). For the Bekenstein--Hawking 
entropy, this gives the standard Bekenstein--Mukhanov value of the area spectrum parameter, which is used as the reference limit. The same discrete 
construction is then applied to generalized entropy functionals. For Barrow entropy, the parameter \(\gamma\) becomes level dependent, while the 
relative separation between adjacent area levels still vanishes for large \(n\). For the modified R\'enyi entropy, the nonsingular branch has 
vanishing relative spacing at large \(n\), whereas the singular branch develops a finite-level pole. For the modified Kaniadakis entropy, the small 
\(\kappa\) expansion shows that a fixed deformation parameter prevents the relative area spacing from vanishing in the large \(n\) limit.
Overall, the results suggest that Landauer's principle provides a useful way to analyze generalized entropic extensions of the Bekenstein--Mukhanov approach.
\end{abstract}

\maketitle

\section{introduction}

Black hole area quantization is rooted in the idea that, for a non-extremal black hole, the horizon area has the character of a 
classical adiabatic invariant~\cite{beken1,beken2,mukha,hod1,hod2,cgs}. As emphasized in Ref.~\cite{hod2}, the Ehrenfest principle~\cite{ehr} 
then suggests that the corresponding quantum observable should have a discrete spectrum. A minimal
implementation of this idea is to assume an evenly spaced area spectrum,
\begin{equation}
\label{area1}
A_n=\gamma \ell_P^2 n \,,\qquad n=1,2,3,\ldots \,,
\end{equation}
where \(\ell_P\) is the Planck length and \(\gamma\) is a dimensionless
parameter.

In the Bekenstein--Mukhanov approach~\cite{bekmuk}, this 
parameter is fixed by the number of microscopic configurations associated with the area levels. If the number of configurations corresponding 
to the area level labeled by \(n\) is taken as \(W=k^n\), then the entropy is
\begin{equation}
S = k_B \ln W = k_B n\ln k \, .
\end{equation}
Matching this expression with the Bekenstein--Hawking entropy,
\begin{equation}
\label{bhal}
S_{BH} = \frac{k_B A}{4\ell_P^2} \,,
\end{equation}
one obtains
\begin{equation}
A_n=4\ell_P^2 n\ln k \,,
\end{equation}
so that \(\gamma=4\ln k\). Therefore, the spacing of the area spectrum is not arbitrary, but is fixed by the underlying microscopic information 
content of the horizon.

In this work, we revisit this problem from a complementary point of view.
Instead of fixing \(\gamma\) through a degeneracy assumption, we use
Landauer's principle~\cite{landauer,bri,land2} as the physical criterion
associated with the erasure of one bit of information. 
For that matter, we retain the linear spectral ansatz
\(A_n=\gamma(n)\ell_P^2 n\), while allowing the coefficient
\(\gamma\) to depend on the level \(n\). Therefore, except in the
Bekenstein--Hawking case, the resulting spectrum is not uniformly
spaced.
The basic prescription then
is to identify the entropy difference between two consecutive area levels with
the corresponding elementary entropy increment,
\begin{equation}
\label{landv}
\Delta S=k_B\ln 2 \, .
\end{equation}
This condition determines the parameter \(\gamma\) for each entropy
functional analyzed below.

We also investigate the macroscopic behavior of the area spectrum through the ratio
\(\Delta A_n/A_n\). Although the spectrum remains discrete at the fundamental
level, this ratio controls the separation between neighboring levels at large
\(n\).
We show that \(\Delta A_n/A_n\) vanishes at large \(n\) for the standard,
Barrow, and modified R\'enyi cases, with the latter restricted to its
nonsingular branch. For the modified Kaniadakis entropy, we work in the
small \(\kappa\) regime and show that, if the deformation parameter is kept
fixed, this ratio does not vanish in the large \(n\) limit. A consistent
macroscopic regime can then be recovered when \(\kappa\) is interpreted as an
effective scale-dependent parameter.

The paper is organized as follows. In Sec.~2, we review the use of
Landauer's principle in the Bekenstein--Hawking case and recover the standard
area spacing. It is worth mentioning that the corresponding value of the
parameter \(\gamma\) was previously obtained in Ref.~\cite{bgs}. In Sec.~3,
we apply the same prescription to the Barrow entropy~\cite{barrow}. In Sec.~4,
we analyze the modified R\'enyi entropy~\cite{renyi} and discuss the regular
and singular branches. In Sec.~5, we consider the modified Kaniadakis
entropy~\cite{aa} in the small \(\kappa\) regime and 
discuss the conditions required for a consistent macroscopic regime.
Finally, Sec.~6 contains our conclusions.

\section{Landauer principle}

Landauer's principle provides a direct link between information theory and thermodynamics \cite{landauer,bri,land2}. It states that erasure of one bit of 
information requires a minimum energy cost, according to the inequality
\begin{eqnarray}
\label{land}
\Delta E \geq k_B T \ln 2 \,,
\end{eqnarray}
where $k_B$ is the Boltzmann constant and $T$ is the temperature. The factor $\ln 2$ reflects the entropy change associated with the loss of a single bit, 
while the corresponding energy is dissipated as heat. This principle has been supported by both experimental and theoretical results \cite{bap,jgb,yxr,rw,ebor,ebor2}, 
and is now regarded as a basic result in the thermodynamics of information.
Landauer's principle has also been applied in several other important contexts in gravitational physics. In particular, it has been argued that Hawking evaporation 
satisfies the Landauer bound in its saturated form \cite{cl}. Related discussions have connected this principle to the Hawking temperature \cite{abreu},  
cosmological horizons \cite{otr,ikr}, the second law of thermodynamics in cosmology \cite{ops}, 
and the Immirzi parameter in Loop Quantum Gravity \cite{euet}.   
Recent reviews on Landauer's principle can be found in \cite{Georgescu2021, Chattopadhyay:2025vis}.

Although a fully general formulation of Landauer's principle in curved
spacetime is not yet established, we adopt its local form for an observer
coupled to a thermal environment. In a static curved spacetime, the temperature entering the local form of Landauer's bound is the
temperature measured by that observer. For a static observer, it obeys
Tolman's relation \cite{Tolman:1930ona}
\begin{equation}
T_{\rm loc}(r)\sqrt{-g_{tt}(r)}=T_\infty \,.
\end{equation}
The corresponding locally measured energy variation is redshifted by the same
factor,
\begin{equation}
\Delta E_{\rm loc}(r) {\sqrt{-g_{tt}(r)}}
=
\Delta E_\infty \,.
\end{equation}
Therefore, the local relation
\begin{equation}
\Delta E_{\rm loc}\geq k_B T_{\rm loc}\ln 2
\end{equation}
is equivalent to
\begin{equation}
\Delta E_\infty\geq k_B T_\infty\ln 2 \,.
\end{equation}
For the Schwarzschild black hole considered below, \(T_\infty\) is the Hawking
temperature \(T\), and the mass variation \(\Delta M\) is the corresponding
energy variation measured at infinity. Adopting a system of units in which $\hbar=c=1$, one has
\begin{eqnarray}
\label{bhe}
S_{BH} = 4 \pi k_B G M^2 \,,
\end{eqnarray}
where $G$ is Newton's constant and $M$ is the black hole mass. Differentiating with respect to $M$, one obtains
\begin{eqnarray}
\label{bht}
\frac{dS_{BH}}{dM}=\frac{1}{T}=8 \pi k_B G M \,.
\end{eqnarray}
For a small mass variation \(\Delta M\), this gives
\begin{eqnarray}
\label{dm}
\Delta S_{BH} = 8 \pi k_B G M \Delta M \, .
\end{eqnarray}
If this entropy change is associated with the erasure of one bit,
Eq.~(\ref{landv}) gives
\begin{eqnarray}
\label{ldm}
\Delta M = \frac{\ln 2}{8 \pi G}\frac{1}{M} \, ,
\end{eqnarray}
which, using Eq.~(\ref{bht}), can be rewritten as
\begin{eqnarray}
\label{lds}
\Delta M = k_B T \ln 2 \, .
\end{eqnarray}
Thus, Eqs.~(\ref{landv}) and (\ref{lds}) show that an elementary entropy variation associated with Hawking evaporation 
is consistent with the saturation of the Landauer bound.

To apply Landauer's principle to the Bekenstein--Hawking area law,
Eq.~(\ref{bhal}), we first rewrite it in terms of the discrete area
spectrum, Eq.~(\ref{area1}), which gives
\begin{equation}
\label{bhan}
S_{BH} = \frac{k_B \,\gamma \, n}{4} \, .
\end{equation}
The exact entropy variation between two consecutive levels is then
\begin{equation}
\label{vs1}
\Delta S_{BH} = S_{BH}(n+1)-S_{BH}(n)
= \frac{k_B \,\gamma}{4} \, .
\end{equation}
Combining Eq.~(\ref{vs1}) with the Landauer prescription,
Eq.~(\ref{landv}), we obtain the expression for the parameter
\(\gamma\) appearing in Eq.~(\ref{area1}),
\begin{eqnarray}
\label{gammaland}
\gamma = 4\ln 2 \, .
\end{eqnarray}
Thus, the Landauer prescription reproduces the Bekenstein--Mukhanov area spacing.

Substituting Eq.~(\ref{gammaland}) into Eq.~(\ref{area1}), the spectrum becomes
\begin{eqnarray}
\label{area1n}
A_n = \left( 4 \, \ell_P^2 \, \ln 2 \right) n \,.
\end{eqnarray}
The corresponding area spacing
\begin{equation}
\label{dana1}
\Delta A_n = 4 \, \ell_P^2 \, \ln 2
\end{equation}
leads to a $1/n$ relative separation between adjacent levels, i.e.,
\begin{equation}
\frac{\Delta A_n}{A_n}
=
\frac{4 \, \ell_P^2 \, \ln 2}{\left( 4 \, \ell_P^2 \, \ln 2 \right)n}
=
\frac{1}{n}\,.
\end{equation}
Consequently, for large $n$, we have
\begin{equation}
\lim_{n\to\infty}\frac{\Delta A_n}{A_n}=0\,.
\end{equation}
Therefore, although the area spectrum remains discrete, the relative spacing between adjacent area levels becomes negligible for 
large \(n\). At macroscopic scales, the relevant behavior is controlled by the vanishing of the relative spacing rather than by the individual 
level structure.

\section{Barrow entropy}
The possibility that quantum gravitational effects give rise to fractal deformations of the black hole horizon was considered by Barrow in \cite{barrow}.
In this framework, the entropy associated with the horizon is modified and takes the form
\begin{eqnarray}
\label{sbarrow}
 S_B = k_B \left( \frac{A}{4 \ell_P^2} \right)^{1+\frac{\Delta}{2}} \,,
\end{eqnarray}
where $A$ is the usual horizon area and $\Delta$ denotes the Barrow deformation parameter \cite{barrow}. This expression differs from the standard 
quantum corrections to black hole entropy, which usually involve logarithmic terms \cite{sari,sari2,aab,kj}, although it formally resembles Tsallis- 
type nonextensive entropy \cite{tsallis,ww,tsallis2,nof}. The exponent in Eq.~(\ref{sbarrow}) characterizes the deviation from the Bekenstein--Hawking area 
law induced by quantum gravitational effects. 
In principle, Barrow proposed a range variation $0 \leq \Delta \leq 1$, in which 
$\Delta=0$ recovers the standard Bekenstein--Hawking 
entropy, while $\Delta=1$ corresponds to the case of maximal deformation \cite{gen1,gen2}.  However, extensions of $\Delta$ to negative values leading to a broader 
range of $-1 < \Delta \leq 1$ have recently been supported both by observational data as well as from theoretical arguments \cite{Anagnostopoulos:2020ctz, Jizba:2024klq, 
Luciano:2025hjn}.

To determine the parameter \(\gamma\) appearing in the area spectrum, Eq.~(\ref{area1}), we express the Barrow entropy as
\begin{equation}
S_B(n)=k_B\left(\frac{\gamma \, n}{4}\right)^{1+\frac{\Delta}{2}} %\,.
\end{equation}
and consider the exact entropy variation between two consecutive area levels
\begin{equation}
\Delta S_B=S_B(n+1)-S_B(n) \,,
\end{equation}
which leads to
\begin{equation}
\Delta S_B
=
k_B\left(\frac{\gamma}{4}\right)^{1+\frac{\Delta}{2}}
\left[(n+1)^{1+\frac{\Delta}{2}}-n^{1+\frac{\Delta}{2}}\right] \,.
\end{equation}
By imposing Landauer's principle, \(\Delta S_B=k_B\ln 2\), we obtain the exact expression
\begin{equation}
\label{gbar}
\gamma_B=
4\left[
\frac{\ln 2}
{(n+1)^{1+\frac{\Delta}{2}}-n^{1+\frac{\Delta}{2}}}
\right]^{\frac{2}{2+\Delta}} \,.
\end{equation}
In the limit \(\Delta \to 0\), Eq.~(\ref{gbar}) recovers the standard Bekenstein--Mukhanov value \(\gamma=4\ln 2\).

Figure 1 shows the parameter \(\gamma_B\), given by Eq.~(\ref{gbar}), as a function of \(n\),
for different values of \(\Delta\). Here, the spectrum starts at \(n=1\), while
\(\Delta\) is restricted to the interval \(0\leq\Delta\leq 1\). One sees that, for
\(\Delta\neq 0\), \(\gamma_B\) decreases as \(n\) increases. This may be interpreted
as an effective running of \(\gamma_B\) induced by the combination of the Barrow
entropy with the Landauer principle.

\begin{figure}[H]
	\centering
	\includegraphics[height=7cm,width=10cm]{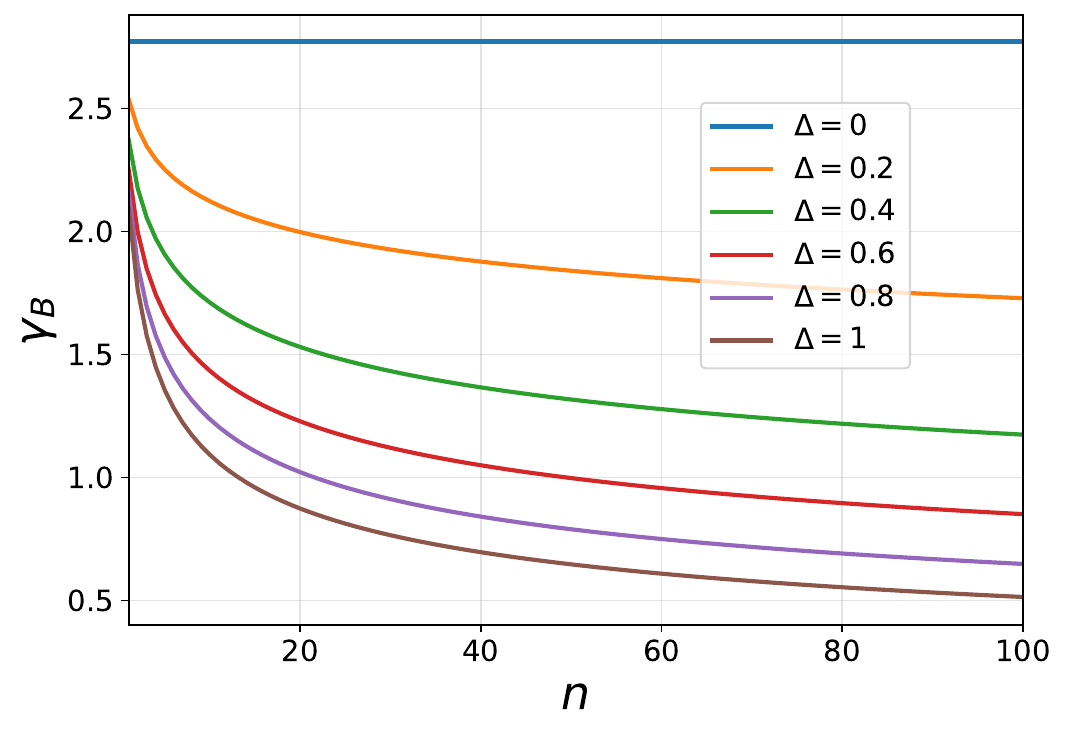}
	\caption{Values of the parameter $\gamma_B$, given by Eq.~(\ref{gbar}), as a function of $n$ for different values of $\Delta$ in the Barrow 
    entropy framework.}
	\label{barrow}
\end{figure}

By substituting the exact expression for the Barrow parameter, Eq.~(\ref{gbar}), into the area spectrum, Eq.~(\ref{area1}), we obtain
\begin{equation}
\label{abar}
A_n =
4\,\ell_P^2\, n
\left[
\frac{\ln 2}
{(n+1)^{1+\frac{\Delta}{2}}-n^{1+\frac{\Delta}{2}}}
\right]^{\frac{2}{2+\Delta}} \,.
\end{equation}
In the large \(n\) regime, Eq.~(\ref{abar}) reduces to
\begin{equation}
\label{abar_large}
A_n \simeq
4\ell_P^2
\left(
\frac{\ln 2}{1+\Delta/2}
\right)^{\frac{2}{2+\Delta}}
n^{\frac{2}{2+\Delta}} \,.
\end{equation}
Therefore, \(A_n\to\infty\) as \(n\to\infty\).

The discrete spacing is defined as
\begin{equation}
\Delta A_n=A_{n+1}-A_n \,.
\end{equation}
Using Eq.~(\ref{abar}), one obtains
\begin{eqnarray}
\label{deltaA_exact_barrow_full}
\Delta A_n
&=&
4\ell_P^2
\Bigg[
(n+1)
\left(
\frac{\ln 2}
{(n+2)^{1+\frac{\Delta}{2}}-(n+1)^{1+\frac{\Delta}{2}}}
\right)^{\frac{2}{2+\Delta}}
\nonumber\\
&&
\hspace{1.0cm}
-
n
\left(
\frac{\ln 2}
{(n+1)^{1+\frac{\Delta}{2}}-n^{1+\frac{\Delta}{2}}}
\right)^{\frac{2}{2+\Delta}}
\Bigg] \,.
\end{eqnarray}
Combining this result with Eq.~(\ref{abar}), the exact relative spacing reads
\begin{eqnarray}
\label{ratio_deltaA_A_barrow_exact}
\frac{\Delta A_n}{A_n}
&=&
\frac{n+1}{n}
\left[
\frac{
(n+1)^{1+\frac{\Delta}{2}}-n^{1+\frac{\Delta}{2}}
}{
(n+2)^{1+\frac{\Delta}{2}}-(n+1)^{1+\frac{\Delta}{2}}
}
\right]^{\frac{2}{2+\Delta}}
-1 \,.
\end{eqnarray}
For \(n\gg 1\), Eq.~(\ref{ratio_deltaA_A_barrow_exact}) gives
\begin{eqnarray}
\frac{\Delta A_n}{A_n}
&\simeq&
\frac{2}{2+\Delta} \,\frac{1}{n} \,.
\label{ratio_deltaA_A_barrow_large_n}
\end{eqnarray}
It follows that
\begin{equation}
\frac{\Delta A_n}{A_n}
\to 0
\qquad
(n\to\infty) \,.
\end{equation}
Thus, the fundamental discreteness of the area spectrum is preserved, whereas the relative spacing between adjacent levels 
vanishes for large \(n\). This is the sense in which the macroscopic limit becomes insensitive to the underlying level structure.

\section{Modified R\'enyi entropy}

To introduce the modified Rényi entropy, we first briefly review Tsallis statistics. 
Tsallis statistics \cite{tsallis,t2,t3} provides a nonextensive extension of the usual 
Boltzmann--Gibbs (BG) framework, in which the standard definition of entropy is generalized 
by means of an entropic index parameter $q$. The corresponding entropy is given by
\begin{eqnarray}
S_q = k_B \, \frac{1 - \sum_{i=1}^{W} p_i^q}{q - 1} \,,
\end{eqnarray}
where $k_B$ is the Boltzmann constant, $p_i$ denotes the probability of the $i$th microstate, 
$q$ is the nonextensivity entropic parameter, and $W$ represents the total number of accessible 
microstates. In the limit $q \to 1$, this expression reduces to the standard BG entropy. 
Although Tsallis entropy preserves the main properties expected from an entropy measure, it is 
not additive in the usual sense. Instead, it satisfies a pseudo additivity relation.
This formalism has found applications in a wide variety of physical contexts, including L\'evy-type anomalous diffusion \cite{t4}, 
turbulence in pure electron plasmas \cite{t5}, position-dependent mass formalisms \cite{daCosta:2018qsp, Borges, Thibes:2025, Thibes:2025b}, gravitational 
systems \cite{t6A, t6B, t6C,t7,t8,t9}, and astrophysical scenarios \cite{Abrahao:2016hnh,piz,jsa,tj}.

In the microcanonical ensemble, for which all microstates are equally probable, the Tsallis entropy takes the simpler form
\begin{eqnarray}
\label{km}
S_q = k_B\,\frac{W^{1-q} - 1}{1-q} \,.
\end{eqnarray}
As expected, Eq.~(\ref{km}) reproduces the BG result $S = k_B \ln W$ when $q \to 1$.

We now assume that the Bekenstein--Hawking entropy \(S_{BH}\), given by
Eq.~(\ref{bhal}), can be described within this nonextensive framework by identifying
it with the microcanonical Tsallis entropy,
\begin{equation}
\label{kbh} 
k_B\,\frac{W^{1-q} - 1}{1-q} = S_{BH} .
\end{equation}
Solving Eq.~(\ref{kbh}) for the number of microstates, one obtains
\begin{eqnarray}
\label{w}
W = \left( 1 + \lambda \frac{S_{BH}}{k_B} \right)^{\frac{1}{\lambda}} \,,
\end{eqnarray}
where we have defined $\lambda \equiv 1-q$. 
Using this expression in the standard BG relation \(S=k_B\ln W\), one obtains
\begin{eqnarray}
\label{rmod}
S_R = \frac{k_B}{\lambda} \ln \left( 1 + \lambda \frac{S_{BH}}{k_B} \right) \,.
\end{eqnarray}
Equation~(\ref{rmod}) defines a deformed R\'enyi entropy \cite{Renyi1960},
in a form previously discussed in the literature 
(see, for example, Refs.~\cite{ci,awn}). In what follows, we refer to this
quantity as the modified R\'enyi entropy (MRE). In the limit
\(\lambda \to 0\), Eq.~(\ref{rmod}) consistently reduces to the usual
Bekenstein--Hawking entropy, \(S_{BH}\).

At this point, it is useful to recall that R\'enyi entropy and related deformations are widely used in quantum information theory as generalized 
measures of uncertainty and entanglement. They also arise naturally in the replica construction of the von Neumann entropy in conformal field 
theory and have been applied to entanglement spectra in systems ranging from black hole horizons to quantum many body systems. See, for instance, 
Refs.~\cite{Puertas-Centeno:2018xht, Zozor:2011qqw, calabrese,nishi}.

To determine the parameter \(\gamma\) in Eq.~(\ref{area1}) within the MRE framework, we rewrite Eq.~(\ref{rmod}) as
\begin{equation}
\label{rmod2}
S_R(n)=\frac{k_B}{\lambda}\ln\left(1+\lambda\frac{\gamma \, n}{4}\right) \, ,
\end{equation}
where Eq.~(\ref{area1}) and the standard relation \(S_{BH}=k_B A/(4\ell_P^2)\) have been used. We then consider the exact entropy variation between two 
consecutive area levels,
\begin{equation}
\Delta S_R=S_R(n+1)-S_R(n) \,,
\end{equation}
which gives
\begin{equation}
\label{drmod_exact}
\Delta S_R
=
\frac{k_B}{\lambda}
\ln\left(
\frac{1+\lambda \frac{\gamma}{4}(n+1)}
     {1+\lambda \frac{\gamma}{4}n}
\right) \,.
\end{equation}
By imposing Landauer's principle, \(\Delta S_R=k_B\ln 2\), we obtain the exact expression
\begin{equation}
\label{gammar}
\gamma_R=
\frac{4\left(2^\lambda-1\right)}
{\lambda\left[1-n\left(2^\lambda-1\right)\right]} \, .
\end{equation}
In the limit \(\lambda \to 0\), Eq.~(\ref{gammar}) consistently recovers the standard Bekenstein--Mukhanov result \(\gamma=4\ln 2\).
From Eq.~(\ref{gammar}), one can directly distinguish the two branches of the solution. For \(\lambda>0\), 
since \(2^\lambda-1>0\), the denominator vanishes at
\begin{equation}
n_c=\frac{1}{2^\lambda-1} \,,
\end{equation}
so that \(\gamma_R\) develops a pole at \(n=n_c\). Moreover, \(\gamma_R\)
becomes negative for \(n>n_c\). Thus, the branch \(\lambda>0\) is singular and is
regular only for \(n<n_c\). By contrast, for \(\lambda<0\), one has
\(2^\lambda-1<0\), and the denominator in Eq.~(\ref{gammar}) remains positive
for all \(n\geq1\). Hence, no divergence arises, and \(\gamma_R\) stays positive
for all \(n\). In this case, \(\gamma_R\to0\) as \(n\to\infty\). Therefore, the
negative-\(\lambda\) branch is free of divergences and remains positive for all
\(n\geq1\), whereas the positive-\(\lambda\) branch does not define a positive
area spectrum beyond the pole.

It should be emphasized that the pole found for the positive-\(\lambda\)
branch does not, by itself, render the MRE unphysical. Instead, within the
present phenomenological construction, it signals an incompatibility among the
assumed area-spectrum ansatz \(A_n=\gamma(n)\ell_P^2 n\), the MRE with fixed
positive \(\lambda\), and the Landauer condition
\(\Delta S=k_B\ln 2\) imposed on consecutive levels. Under these simultaneous
assumptions, the coefficient \(\gamma_R(n)\) develops a pole at \(n=n_c\) and
becomes negative beyond it. Consequently, the resulting spectrum cannot be
continued as a positive area spectrum to arbitrarily large \(n\).

Whether this obstruction reflects a physical inconsistency of the MRE in the
positive-\(\lambda\) branch or a limitation of the simultaneous use of the
Landauer condition and the assumed area-spectrum parametrization cannot be
decided within the present analysis. Resolving this issue would require a
microscopic derivation of the black hole horizon area spectrum. Therefore, we
do not interpret the pole as establishing the physical exclusion of the
positive-\(\lambda\) branch.

Figure~2 presents $\gamma_R$ as a function of $n$ for different negative values of $\lambda$. As $n$ increases, $\gamma_R$ decreases toward zero, 
indicating an effective running of $\gamma_R$ induced by the Landauer criterion.

\begin{figure}[H]
	\centering
	\includegraphics[height=7cm,width=10cm]{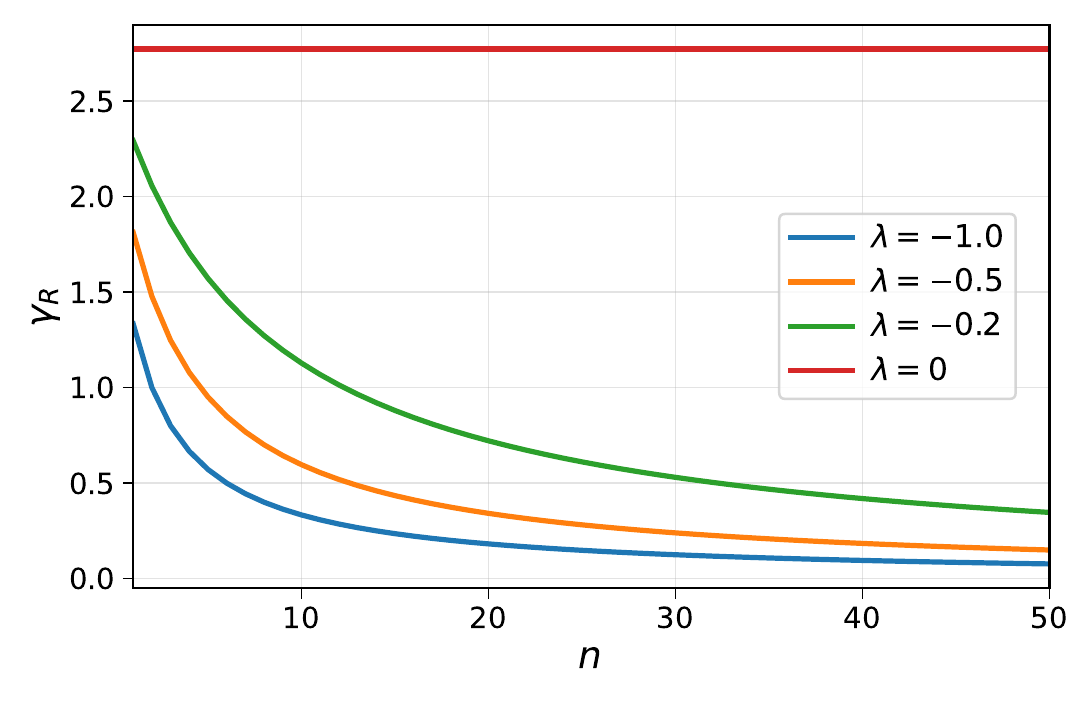}
	\caption{Parameter \(\gamma_R\), given by Eq.~(\ref{gammar}), as a function of \(n\)
for selected values of \(\lambda \leq 0\) in the MRE framework.}
	\label{renyi}
\end{figure}

Substituting Eq.~(\ref{gammar}) into Eq.~(\ref{area1}), for the regular negative-\(\lambda\) branch within the present 
construction, \(\lambda<0\), yields
\begin{equation}
\label{arearenyi2}
A_n=
\frac{4\,\ell_P^2\,n\left(1-2^\lambda\right)}
{|\lambda|\left[1+n\left(1-2^\lambda\right)\right]} \, ,
\end{equation}
which is positive and finite for all \(n\geq 1\). In the large \(n\) limit, one obtains
\begin{equation}
\label{arearenyi_limit}
A_n \longrightarrow \frac{4\,\ell_P^2}{|\lambda|}
\qquad
(n\to\infty) \, .
\end{equation}
Thus, for \(\lambda<0\), the area approaches a finite asymptotic value. From Eq.~(\ref{arearenyi2}), the discrete spacing is
\begin{equation}
\label{deltaarenyi}
\Delta A_n
=
A_{n+1}-A_n
=
\frac{4\,\ell_P^2\left(1-2^\lambda\right)}
{|\lambda|
\left[1+n\left(1-2^\lambda\right)\right]
\left[1+(n+1)\left(1-2^\lambda\right)\right]} \, .
\end{equation}
Dividing Eq.~(\ref{deltaarenyi}) by Eq.~(\ref{arearenyi2}), one obtains
\begin{equation}
\label{ratioarenyi}
\frac{\Delta A_n}{A_n}
=
\frac{1}
{n\left[1+(n+1)\left(1-2^\lambda\right)\right]} \, .
\end{equation}
For fixed \(\lambda<0\), the large \(n\) behavior is
\begin{equation}
\label{ratioarenyi_large}
\frac{\Delta A_n}{A_n}
\simeq
\frac{1}{n^2\left(1-2^\lambda\right)}
\qquad
(n\gg 1) \, .
\end{equation}
Hence, the relative spacing vanishes in the macroscopic limit,
\(\Delta A_n/A_n\to 0\) as \(n\to\infty\).
For \(\lambda<0\), the modified R\'enyi area spectrum is free of divergences and approaches a finite limiting value. 
At the same time, the relative spacing between adjacent levels vanishes for large \(n\). This behavior contrasts with the Barrow 
entropy case, where the area spectrum grows without bound as \(n\to\infty\).

The finite limiting value found in the modified R\'enyi case
deserves some further comment. For \(\lambda<0\), the area spectrum develops
an accumulation point at \(A_{\max}=4\ell_P^2/|\lambda|\), namely,
\(A_n\to A_{\max}\) while \(\Delta A_n/A_n\to0\). Although the spectrum
remains discrete, the adjacent levels become increasingly dense as they
approach \(A_{\max}\). In fact, Eq.~(\ref{deltaarenyi}) shows that the level
spacing vanishes as \(n\to\infty\). Hence, no macroscopic discreteness survives
in the large-\(n\) regime.

This bounded-area behavior should not be interpreted as a physical prediction
of the MRE alone. It follows from the simultaneous assumptions of a fixed
negative deformation parameter in the MRE and the Landauer-based linear
spectral ansatz. Its origin may lie in the MRE for the negative-\(\lambda\)
branch, in the Landauer-based spectral construction at arbitrarily large
\(n\), or in both. Distinguishing among these possibilities would require a
microscopic derivation of the entropy functional and the horizon area spectrum.

\section{Modified Kaniadakis entropy}

Kaniadakis statistics defines a nonextensive extension of the standard
Boltzmann--Gibbs (BG) framework through a real deformation parameter
\(\kappa\)~\cite{k1,k2,k3,k4}. In this formalism, the entropy is written as
\begin{eqnarray}
S_\kappa = -k_B \sum_{i=1}^W
\frac{p_i^{1+\kappa}-p_i^{1-\kappa}}{2\kappa}\,,
\end{eqnarray}
where \(p_i\) denotes the probability of the \(i\)th microstate and \(W\) is
the total number of accessible states. The standard BG entropy is recovered
in the limit \(\kappa\to0\). Owing to its deformed structure, this entropy has
found applications in several areas, including relativistic systems,
gravitation, and cosmology~\cite{ks,kqs,aabn,aamp,lbs,ggl, Santos:2022fbq,as, daCosta:2020mbf, Santos:2020txg}. A homotopic
transformation connecting Kaniadakis and Tsallis entropies has recently been
investigated in Refs.~\cite{Thibes:2025,Thibes:2025c}.

In the microcanonical case, where all microstates are equally probable, the
above expression reduces to
\begin{eqnarray}
S_\kappa = k_B \frac{W^\kappa-W^{-\kappa}}{2\kappa}\,.
\end{eqnarray}
This expression reduces to \(S=k_B\ln W\) for \(\kappa\to0\), and the
deformation parameter is usually restricted to \(0\leq\kappa<1\).

In order to construct an entropy analogous to the MRE within the Kaniadakis framework,
the authors of Ref.~\cite{aa} proposed a modified Kaniadakis entropy as follows.
They identified the microcanonical Kaniadakis entropy with the Bekenstein--Hawking
entropy, Eq.~(\ref{bhal}), namely
\begin{equation}
k_B \frac{W^\kappa-W^{-\kappa}}{2\kappa}=S_{BH} \,.
\end{equation}
Solving for \(W\), one obtains
\begin{equation}
W=\left(
\kappa\frac{S_{BH}}{k_B}
+\sqrt{1+\kappa^2\frac{S_{BH}^2}{k_B^2}}
\right)^{1/\kappa} \,.
\end{equation}
Substituting this result into the relation \(S = k_B \ln W\) yields the modified
Kaniadakis entropy (MKE)
\begin{equation}
\label{mkeeq}
S_\kappa^*=
\frac{k_B}{\kappa}
\ln\left(
\kappa\frac{S_{BH}}{k_B}
+\sqrt{1+\kappa^2\frac{S_{BH}^2}{k_B^2}}
\right) \,.
\end{equation}
In the limit \(\kappa \to 0\), this expression reduces to the standard
Bekenstein--Hawking entropy, \(S_{BH}\). The MKE may
therefore be regarded as a relativistic analogue of R\'enyi-type constructions.
The thermodynamics induced by this entropy differs from that of the Bekenstein--Hawking case. 
In particular, the associated Hawking temperature may lead to a positive heat capacity for Schwarzschild black holes. 
As a result, above a critical mass, the temperature increases with the black hole mass. 
The corresponding Hawking mass loss rate is also found to be equal to or larger than the standard one, leading to a shorter evaporation time. 
This feature may be relevant for discussions of the non-observation of primordial black holes. 
Since black hole evaporation is a quantum effect, these results suggest that the MKE provides an effective parametrization of quantum corrections to black hole 
thermodynamics. Further details can be found in Refs.~\cite{aa,aaacnt}.

In order to determine the parameter $\gamma$ within the MKE framework, we first rewrite
Eq.~(\ref{mkeeq}) in terms of the discrete area spectrum given in Eq.~(\ref{area1}). This gives
\begin{equation}
\label{mken}
S_\kappa^*(n)
=
\frac{k_B}{\kappa}
\ln\left[
\kappa\frac{\gamma n}{4}
+
\sqrt{
1+\kappa^2\frac{\gamma^2 n^2}{16}
}
\right].
\end{equation}
At this point, the exact discrete implementation of Landauer's prescription leads to algebraically involved expressions. Since the physically 
relevant regime must recover the Bekenstein--Hawking result as $\kappa\to0$, we shall focus on the small $\kappa$ expansion. This approximation 
is sufficient to identify how the Kaniadakis deformation modifies the area-spectrum parameter and to determine
whether the relative spacing vanishes in the macroscopic regime.
Expanding Eq.~(\ref{mken}) up to order $\kappa^2$, one obtains
\begin{equation}
\label{skexp}
S_\kappa^*(n)
\simeq
k_B
\left[
\frac{\gamma n}{4}
-
\frac{\kappa^2\gamma^3 n^3}{384}
\right] \, .
\end{equation}
The corresponding discrete entropy variation is
\begin{equation}
\label{dskexp}
\Delta S_\kappa^*(n)
=
S_\kappa^*(n+1)-S_\kappa^*(n) \,,
\end{equation}
which gives
\begin{equation}
\label{dskexp2}
\Delta S_\kappa^*(n)
\simeq
k_B
\left[
\frac{\gamma}{4}
-
\frac{\kappa^2\gamma^3}{384}
\left(3n^2+3n+1\right)
\right] \, .
\end{equation}
By imposing Landauer's principle, $\Delta S_\kappa^*(n)=k_B\ln 2$, we obtain
\begin{equation}
\label{gammakeq}
\frac{\gamma}{4}
-
\frac{\kappa^2\gamma^3}{384}
\left(3n^2+3n+1\right)
=
\ln 2 \, .
\end{equation}
Solving this equation perturbatively up to order $\kappa^2$, by writing
\begin{equation}
\gamma_\kappa(n)
=
\gamma_0+\kappa^2\gamma_2+O(\kappa^4) \,,
\end{equation}
one finds
\begin{equation}
\gamma_0=4\ln 2
\end{equation}
and
\begin{equation}
\gamma_2
=
\frac{2}{3}
\left(3n^2+3n+1\right)
(\ln 2)^3 \,.
\end{equation}
Thus, the Landauer selected parameter in the MKE framework is
\begin{equation}
\label{gammaksmall}
\gamma_\kappa(n)
\simeq
4\ln 2
\left[
1+
\frac{\kappa^2(\ln 2)^2}{6}
\left(3n^2+3n+1\right)
\right] \,.
\end{equation}
In the limit $\kappa\to 0$, the standard BG value
$\gamma=4\ln 2$ is recovered.
Substituting Eq.~(\ref{gammaksmall}) into the area spectrum given in Eq.~(\ref{area1}), one obtains
\begin{equation}
\label{aksmall}
A_n
\simeq
4\ell_P^2 n\ln 2
\left[
1+
\frac{\kappa^2(\ln 2)^2}{6}
\left(3n^2+3n+1\right)
\right] .
\end{equation}
The corresponding discrete area spacing is
\begin{equation}
\label{daksmall}
\Delta A_n
=
A_{n+1}-A_n \,.
\end{equation}
Using Eq.~(\ref{aksmall}), one finds
\begin{equation}
\Delta A_n
\simeq
4\ell_P^2\ln 2
\left[
1+
\frac{\kappa^2(\ln 2)^2}{6}
\left(9n^2+15n+7\right)
\right] .
\end{equation}
Therefore, the relative area spacing is
\begin{equation}
\label{ratioaksmall1}
\frac{\Delta A_n}{A_n}
\simeq
\frac{
1+
\frac{\kappa^2(\ln 2)^2}{6}
\left(9n^2+15n+7\right)
}
{
n
\left[
1+
\frac{\kappa^2(\ln 2)^2}{6}
\left(3n^2+3n+1\right)
\right]
} .
\end{equation}
Expanding this expression consistently up to order \(\kappa^2\), one obtains
\begin{equation}
\label{ratioaksmall2}
\frac{\Delta A_n}{A_n}
\simeq
\frac{1}{n}
\left[
1+
\kappa^2(\ln 2)^2
\left(n^2+2n+1\right)
\right] .
\end{equation}
Equivalently,
\begin{equation}
\label{ratioaksmall3}
\frac{\Delta A_n}{A_n}
\simeq
\frac{1}{n}
+
\kappa^2(\ln 2)^2
\left(
n+2+\frac{1}{n}
\right) .
\end{equation}
For large \(n\), the dominant behavior is
\begin{equation}
\label{ratioaksmall4}
\frac{\Delta A_n}{A_n}
\simeq
\frac{1}{n}
+
\kappa^2(\ln 2)^2 n \,.
\end{equation}
Therefore, for fixed $\kappa$, the second term prevents the relative spacing from vanishing in the large $n$ limit.
One possible way to make the relative spacing vanish at large \(n\) is to allow
the deformation parameter to be interpreted as an effective scale-dependent quantity,
\(\kappa=\kappa(n)\), decreasing with \(n\) in such a way that
\begin{equation}
\label{condkappa}
\kappa(n)\sqrt{n}\ll 1
\qquad
(n\gg1) \,.
\end{equation}
This condition is equivalent to $\kappa^2(n)n\ll1$ and ensures that $\Delta A_n/A_n\to0$ as $n\to\infty$.
This asymptotic interpretation is restricted to the regime in which the small-\(\kappa\) expansion remains controlled. 
Thus, in the MKE framework, a fixed deformation parameter does not make the relative spacing vanish in the large \(n\) regime. 
This difficulty can be avoided, for example, by treating \(\kappa\) as an effective scale-dependent parameter rather than as a fixed 
fundamental constant.

The effective dependence \(\kappa=\kappa(n)\) requires some clarification.
This scale-dependent interpretation of \(\kappa\) is analogous to a
renormalization-group flow. Similar scale-dependent ideas have been
considered in other contexts, for example through a running Barrow
parameter~\cite{gen2} and through a possible connection between entropy
corrections and renormalization-group scaling of the entropy
functional~\cite{Ong:2026iwh}. In the present work, however, the dependence
\(\kappa=\kappa(n)\) is introduced solely as an effective parametrization
required to make the relative spacing \(\Delta A_n/A_n\) vanish in the
macroscopic regime. We do not derive an underlying flow equation or identify a
fundamental renormalization scale. A more rigorous connection with
renormalization-group methods lies beyond the scope of the present effective
description.

\section{Conclusions}

In this paper, we examined the implications of Landauer's principle for the quantization
of black hole horizon area in different entropic frameworks. The analysis was performed at
the exact discrete level by identifying the entropy variation between two neighboring area
levels with the elementary one-bit entropy increment. This procedure provides a direct
way to determine the parameter \(\gamma\) appearing in the area spectrum without assuming,
from the beginning, a specific microscopic degeneracy for the horizon states.

For Barrow entropy, the coefficient \(\gamma_B\) becomes dependent on the area level
\(n\) and on the deformation parameter \(\Delta\). The resulting spectrum is no longer
uniformly spaced. Nevertheless, the relative separation between neighboring levels
vanishes in the large \(n\) regime. Therefore, the fundamental discreteness of the spectrum
is preserved, although the separation between consecutive levels becomes negligible at
macroscopic scales.

For the modified R\'enyi entropy, the Landauer prescription separates two branches.
For \(\lambda>0\), the coefficient \(\gamma_R\) itself develops a finite-level pole and 
becomes negative beyond it, so that this branch does not define a positive area spectrum 
past the pole. By contrast, for \(\lambda<0\), no divergence occurs, \(\gamma_R\) remains 
positive for all \(n\), and the area spectrum approaches a finite asymptotic value. 
In this branch, the relative area spacing also vanishes as \(n\to\infty\), so that the 
macroscopic regime becomes insensitive to the underlying level structure.

For the modified Kaniadakis entropy, the small-\(\kappa\) analysis shows that
a fixed deformation parameter prevents the relative area spacing from
vanishing in the large-\(n\) regime. A vanishing relative spacing can instead
be recovered if \(\kappa\) is interpreted as an effective scale-dependent
quantity satisfying \(\kappa^2(n)n\ll 1\).

Taken together, our results show that Landauer's principle can be used beyond the
recovery of the Bekenstein--Mukhanov result. It also provides a useful framework for
analyzing generalized entropic extensions, distinguishing regular branches from singular
ones and clarifying the behavior of the level spacing in the large \(n\) regime.

\section*{ACKNOWLEDGMENTS}
We thank the Referee for a careful reading of the manuscript and constructive comments.
Jorge Ananias Neto acknowledges partial financial support from CNPq, the Brazilian
federal agency for scientific support, through CNPq-PQ Grant No. 305984/2023-3.

\end{document}